\documentclass[twocolumn,showpacs,pre,superscriptaddress,floatfix]{revtex4}

\usepackage{color} 
\usepackage{tabularx} 
\usepackage{epsfig}
\usepackage{amsmath} 
\usepackage{amssymb} 
\usepackage{graphicx}
\usepackage{wasysym}

\begin{document}

\title{Evidence of a glass transition in a 10-state nonmean-field
Potts glass}

\author{Ruben S.~Andrist}
\affiliation {Theoretische Physik, ETH Zurich, CH-8093 Zurich, Switzerland}

\author{Derek Larson}
\affiliation{Department of Physics, National Taiwan University,
Taipei, Taiwan}

\author{Helmut G.~Katzgraber} 
\affiliation {Department of Physics and Astronomy, Texas A\&M University,
College Station, Texas 77843-4242, USA}
\affiliation {Theoretische Physik, ETH Zurich, CH-8093 Zurich, Switzerland}

\date{\today}

\begin{abstract}

Potts glasses are prototype models that have been used to understand
the structural glass transition.  However, in finite space dimensions a
glass transition remains to be detected in the 10-state Potts glass.
Using a one-dimensional model with long-range power-law interactions
we present evidence that a glass transition below the upper
critical dimension can exist for short-range systems at low enough
temperatures. Gaining insights into the structural glass transition
for short-range systems using spin models is thus potentially possible,
yet difficult.

\end{abstract} 

\maketitle

\section{Introduction}

Due to their fascinating properties such as aging, memory
effects and ergodicity-breaking transitions, as well as industrial
applications, structural glasses, supercooled liquids and polymers
have received considerable attention recently.  In particular,
when the temperature is decreased, they undergo a dynamic transition
\cite{goetze:92,angell:95,binder:02} below which the particle-density
correlation length does not decay to zero in the long-time limit
and the evolution becomes nonergodic. However, this transition
is not associated with any thermodynamic singularity.  Hence the
system ``freezes'' in a portion of phase space.  There is a second
transition at a lower temperature \cite{kauzmann:48,gibbs:57} which
can be associated with a thermodynamic singularity and which can be
related to a possible ideal glass transition. Despite ongoing efforts,
the structural glass transition remains to be fully understood.

The $p$-state Potts glass
\cite{elderfield:83,gross:85,carmesin:88,scheucher:90,schreider:95,dillmann:98}
is one of the most versatile models in statistical physics: For
$p = 2$ states it reduces to the well-known Edwards-Anderson
Ising spin glass \cite{edwards:75}, a workhorse in the study
of disordered magnetic systems. For $p = 3$ it can be used to
model orientational glasses \cite{binder:92}, while for $p = 4$
the Potts glass can be used to model quadrupolar glasses. For
large $p > 4$ and no disorder the model shows a first-order
transition.  In particular, infinite-range Potts glasses with
$p>4$ exhibit a transition from ergodic to nonergodic behavior
\cite{elderfield:83,gross:85,carmesin:88,scheucher:90,schreider:95,dillmann:98},
as well as an additional static transition at a lower
temperature. In fact, the equations describing the system's
dynamics near the transition are mathematically related
\cite{kirkpatrick:87b,kirkpatrick:88,kirkpatrick:89} to the
equations of mode-coupling theory, which describe the behavior
found in structural glasses and supercooled liquids. Therefore,
studying the Potts glass with large $p$ could provide, in principle,
some insights into the mechanisms governing the structural glass
{\em transition}.  However, this beneficial relationship seems
to only work when the model is infinite ranged \cite{kob:00}. The
existence of a transition in finite-dimensional systems remains to be
proven \cite{brangian:03,lee:06}. Not only are hypercubic lattices
with large space dimension hard to study numerically, recent work
\cite{cruz:09-ea,alvarez:10-ea} suggests that if there is a transition
for large $p$ it would occur at very low temperatures.

In this work we simulate the 10-states Potts glass on a one-dimensional
ring topology with power-law interactions.  This allows us to
effectively tune the range of the interactions and therefore
the (effective) space dimension for large linear system sizes.
Our results suggest that 10-state Potts glasses should have a very
low finite-temperature transition for finite space dimensions.

The paper is structured as follows. In Sec.~\ref{sec:model} we
introduce the model and observables. Furthermore, we outline the
details of the numerical simulations.  Section \ref{sec:results}
summarizes our findings, followed by concluding remarks.

\section{Model and Observables}
\label{sec:model}

We study a one-dimensional Potts glass with long-range power-law
interactions \cite{kotliar:83,katzgraber:03} and Hamiltonian
${\mathcal H} = -\sum_{i,j} J_{ij} \delta_{q_i,q_j}$, where $q_i \in
\{1,\ldots,10\}$ are $10$-state Potts spins on a ring of length $L$ to
enforce periodic boundary conditions and $\delta_{x,y} = 1$ if $x=y$
and zero otherwise. The sum is over all spins and the interactions
$J_{ij}$ are given by $J_{ij} = \varepsilon_{ij}/r_{ij}^\sigma$, where
$\varepsilon_{ij}$ are Normal distributed with mean $J_0$ and standard
deviation unity. $r_{ij} = (L/\pi)\sin[(\pi |i - j|)/L]$ represents the
geometric distance between the spins on the ring.  For the simulations
we express the Potts glass Hamiltonian using the simplex representation
where the $10$ states of the Potts spins are mapped to the corners of
a hypertetrahedron in nine space dimensions. The state of each spin is
therefore represented by a nine-dimensional unit vector $\vec{S}_i$
taking one of the $10$ possible values satisfying the condition $
\vec{S}^\mu \cdot \vec{S}^{\nu} = [p/(p-1)](\delta_{\mu,\nu} - 1)$
with $\{\mu,\nu\} \in \{1,2, \ldots, 10\}.$ In this representation
the Potts glass Hamiltonian is given by ${\mathcal H} = -\sum_{i,
j} \tilde{J}_{ij} \vec{S}_i \cdot \vec{S}_j$ with $\tilde{J}_{ij} =
J_{ij}(p-1)/p$. In the limit when $\sigma \to 0$, when the system is
infinite ranged (Sherrington-Kirkpatrick limit), we obtain $T_{c}(\sigma
= 0)=1/(p-1)$.

The merit of the long-range one-dimensional model lies in emulating
a short-range topology of varying dimensionality, depending on the
power-law exponent: For $\sigma\le2/3$ the model is in the mean-field
long-range 10-state Potts universality class and, in particular for
$\sigma \le 1/2$ in the infinite-range universality class. However,
for $2/3 < \sigma < 1$ the model is in a nonmean-field universality
class with a finite transition temperature $T_c$.  It can be shown
\cite{kotliar:83} that $\sigma = 2/3$ corresponds exactly to six
space dimensions for a hypercubic lattice. Therefore, $\sigma$
values between $1/2$ and $2/3$ allow us to effectively study
\cite{kotliar:83,katzgraber:03} a short-range hypercubic Potts
glass {\em above} the upper critical dimension $d_{\rm u} = 6$,
whereas when $\sigma > 2/3$ we effectively study a model with a
space dimension {\em below} six dimensions. Thus, by studying the
one-dimensional model we can infer if a transition should be present
for the corresponding short-range hypercubic Potts glass.

The presence of a transition is probed by studying the two-point
finite-size correlation length \cite{palassini:99b}. We measure the
wave-vector-dependent spin-glass susceptibility \cite{katzgraber:09b}
\begin{equation}
\chi_{\rm SG}({\bf k}) = N \sum_{\mu,\nu} [\langle \left|q^{\mu\nu}({\bf
k})\right|^2 \rangle ]_{\rm av}\,,
\label{eq:chi}
\end{equation}
where $\langle \cdots \rangle$ denotes a thermal average,
$[\cdots]_{\rm av}$ an average over the disorder and
\begin{equation}
q^{\mu\nu}({\bf k}) = \frac{1}{N} \sum_i S_i^{\mu(\alpha)} S_i^{\nu(\beta)}
e^{i {\bf k} \cdot {\bf R}_i}\,,
\end{equation}
is the spin-glass order parameter computed over two replicas $(\alpha)$
and $(\beta)$ with the same disorder. The two-point finite-size correlation
length is then given by
\begin{equation}
\xi_L = \frac{1}{2 \sin (k_\mathrm{min}/2)}
\left[\frac{\chi_{\rm SG}({\bf 0})}{\chi_{\rm SG}({\bf k}_\mathrm{min})} 
- 1\right]^{1/(2\sigma -1)} \, ,
\end{equation}
where ${\bf k}_\mathrm{min} = 2\pi/L$ is the smallest nonzero wave
vector. According to finite-size scaling \cite{katzgraber:09b}
\begin{subequations}
\label{eq:xiscale}
\begin{align}
{\xi_L/L^{\nu/3}} &= {\mathcal X} [ L^{1/3} (T - T_c) ] 
\;\;\; (1/2 <\sigma \le 2/3)\, ,
\label{eq:xiscaleMF}
\\
{\xi_L/L} &= {\mathcal X} [ L^{1/\nu} (T - T_c) ]  
\;\;\; 
(2/3 < \sigma)\, ,
\label{eq:xiscaleNMF}
\end{align}
\end{subequations}
where $\nu$ is the critical exponent for the correlation length
and $T_c$ the critical temperature. For $\sigma < 2/3$, $\nu =
1/(2\sigma -1)$. 

In practice, there are corrections to scaling to
Eqs.~(\ref{eq:xiscale}) and so data for different system sizes do
not cross exactly at one point as implied by the finite-size scaling
expressions. The crossings between pairs of system sizes $L$ and $2L$
shift with temperature and tend to a constant for $L \to\infty$. 
In general, $T_c^* =
T_c^\infty + b/L^{\theta}$ with $\theta = 1/\nu + \omega$.  Here we find
empirically that $1/\nu + \omega \approx 1$. We fit $T_c^*(L,2L)$ with
high probability to a linear function in $1/L$. The intercept with the
vertical axis after the fit determines a lower bound for the transition
temperature. Error bars are determined via a bootstrap analysis.

To obtain a better understanding of the corrections to scaling we also
measure the spin-glass susceptibility [Eq.~(\ref{eq:chi}) with ${\bf
k} = 0$].  The finite-size scaling of the spin-glass susceptibility
$\chi_{\rm SG}$ is given by
\begin{subequations}
\label{eq:chiscale}
\begin{align}
{\chi_{\rm SG}/L^{1/3}} &= {\mathcal C} [ L^{1/3} (T - T_c) ] 
\;\;\; (1/2 <\sigma \le 2/3)\, ,
\label{eq:chiscaleMF}
\\
{\chi_{\rm SG}/L^{2 - \eta}} &= {\mathcal C} [ L^{1/\nu} (T - T_c) ] 
\;\;\; 
(2/3 < \sigma)\, .
\label{eq:chiscaleNMF}
\end{align}
\end{subequations}
In general, the exponent $\eta$ has to be known {\em a priori}
to precisely determine the location of $T_c$. However, for the
one-dimensional model $2 - \eta = 2\sigma - 1$ for $\sigma > 2/3$
exactly and so ${\chi_{\rm SG}/L^{2 - \eta}}$ can be treated as a
dimensionless quantity similar to the two-point correlation length.

To prevent ferromagnetic order \cite{gross:85,elderfield:83} we set
the mean of the random interactions to $J_0 = -1$ \cite{brangian:03}
in our simulations. This suppresses the ferromagnetic susceptibility
$\chi_{m} = N \sum_{\mu} [\langle \left|m^{\mu}\right|^2 \rangle]_{\rm
av}$ [$m^{\mu} = (1/N) \sum_i S_i^{\mu}$]. We discuss the case where
$J_0 = -1$ in more detail below.

The simulations are done using the parallel tempering Monte Carlo
technique \cite{hukushima:96}; simulation parameters are shown
in Table~\ref{tab:simparams}.  Equilibration is tested by using an
exact relationship between the energy and four-spin correlators (link
overlap) \cite{katzgraber:01} when the bond disorder is Gaussian,
suitably generalized to Potts spins \cite{lee:06} on a one-dimensional
topology \cite{katzgraber:05c}.

\begin{table}[!tb]
\vspace*{-.21cm}
\caption{
Parameters of the simulations for different exponents $\sigma$. $N_{\rm
sa}$ is the number of samples, $N_{\rm sw}$ is the total number of
Monte Carlo sweeps, $T_{\rm min}$ is the lowest temperature simulated,
and $N_T$ is the number of temperatures used in the parallel tempering
method for each system size $L$.
\vspace*{.1cm}
\label{tab:simparams}}
{\footnotesize
\begin{tabular*}{\columnwidth}{@{\extracolsep{\fill}} c r r r r r}
\hline
\hline
$\sigma$ & $L$ & $N_{\rm sa}$ & $N_{\rm sw}$ & $T_{\rm min}$ & $N_{T}$
\\ 
\hline
$0.60$ & $ 32,48,64,96 $ &    $4000$ & $2^{20}$ & $0.054$ & $41$ \\
$0.60$ & $ 128,192 $ &        $2400$ & $2^{21}$ & $0.054$ & $41$ \\
$0.60$ & $ 256 $ &             $500$ & $2^{22}$ & $0.054$ & $41$ \\
$0.60$ & $ 512 $ &             $200$ & $2^{22}$ & $0.054$ & $41$ \\[1mm]
           
$0.75$ & $ 32,48,64,96 $ &    $4000$ & $2^{20}$ & $0.030$ & $41$ \\
$0.75$ & $ 128 $ &            $1600$ & $2^{22}$ & $0.030$ & $41$ \\
$0.75$ & $ 192 $ &            $1600$ & $2^{24}$ & $0.030$ & $41$ \\
$0.75$ & $ 256 $ &             $500$ & $2^{26}$ & $0.030$ & $41$ \\[1mm]

$0.85$ & $ 32,48,64,96 $ &    $4000$ & $2^{20}$ & $0.018$ & $61$ \\
$0.85$ & $ 128 $ &            $1600$ & $2^{22}$ & $0.018$ & $61$ \\
$0.85$ & $ 192 $ &            $1600$ & $2^{24}$ & $0.025$ & $41$ \\
$0.85$ & $ 256 $ &             $500$ & $2^{26}$ & $0.025$ & $41$ \\
\hline
\hline
\end{tabular*}
\vspace{-.2cm}
}
\end{table}

\begin{figure*}
\centering
\includegraphics[width=0.95\columnwidth]{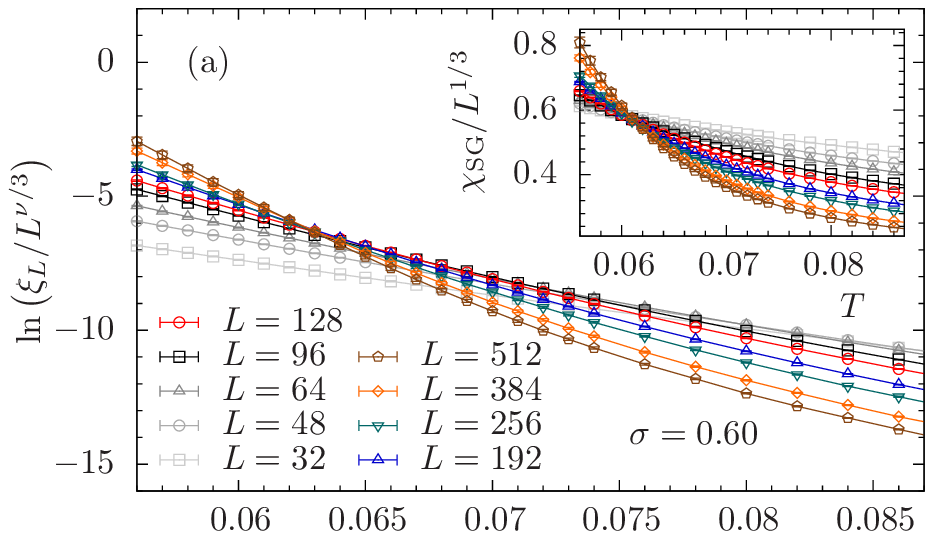}
\includegraphics[width=0.95\columnwidth]{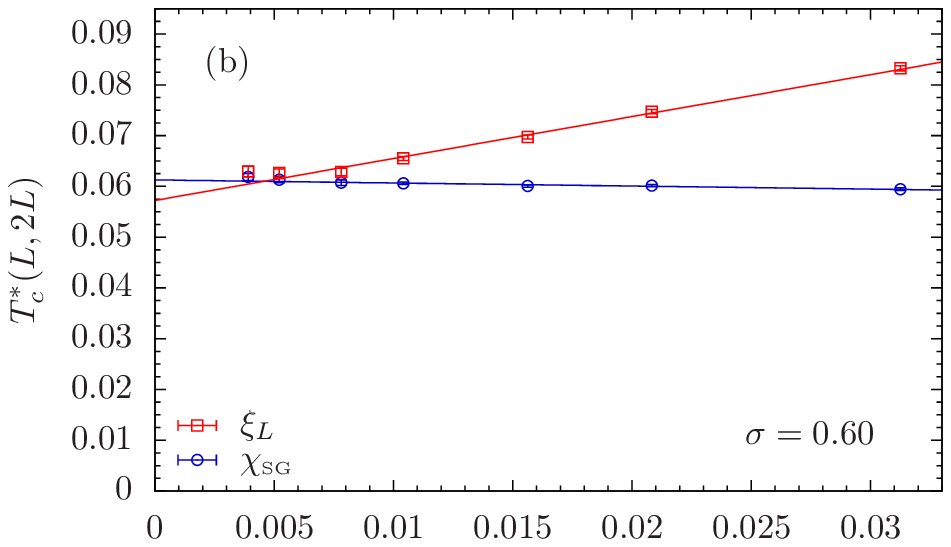}
\includegraphics[width=0.95\columnwidth]{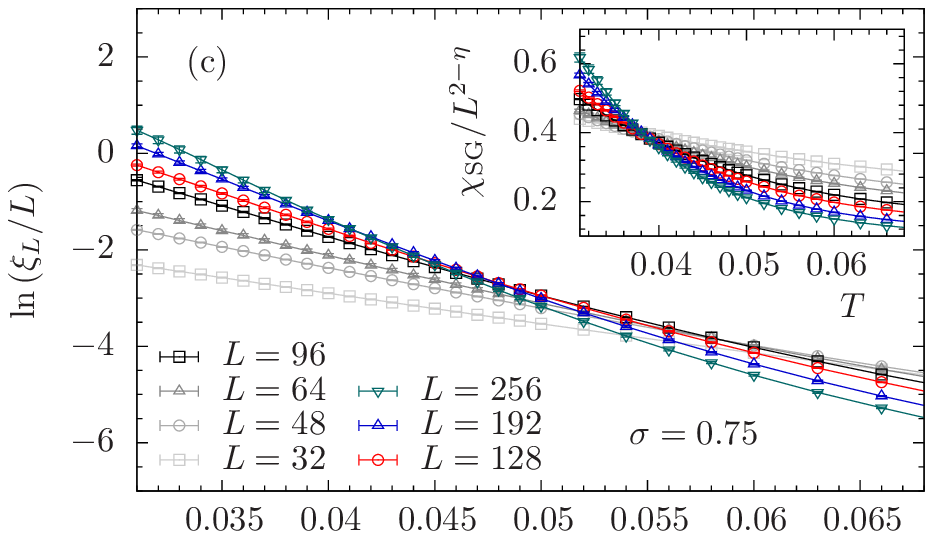}
\includegraphics[width=0.95\columnwidth]{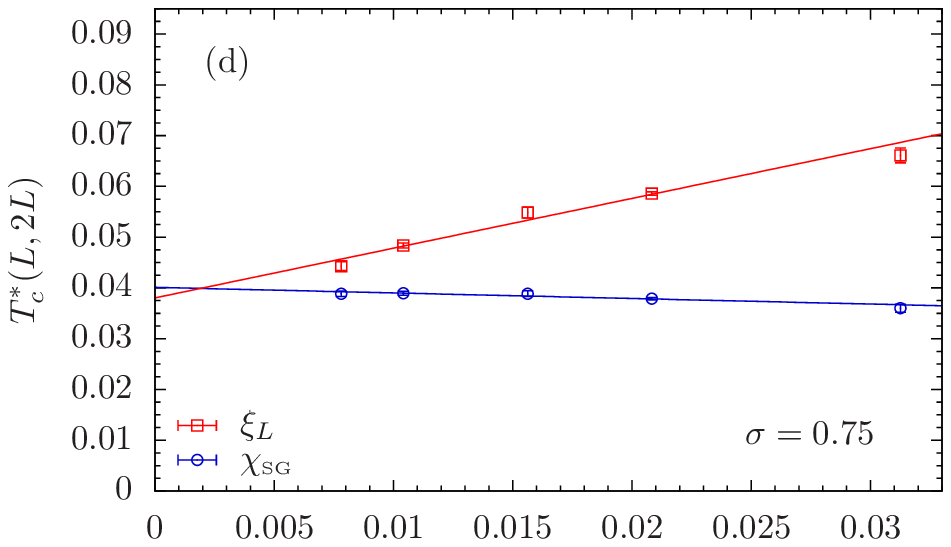}
\includegraphics[width=0.95\columnwidth]{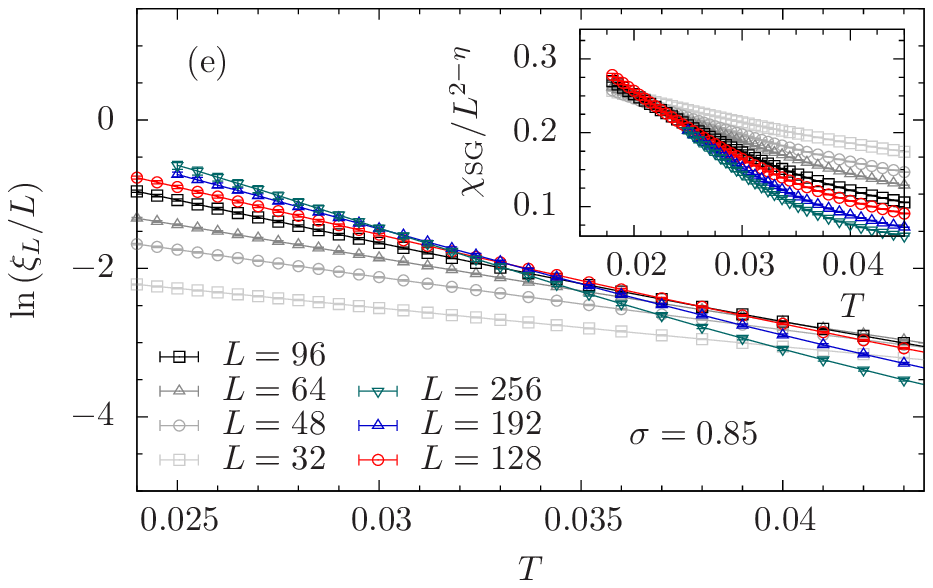}
\includegraphics[width=0.95\columnwidth]{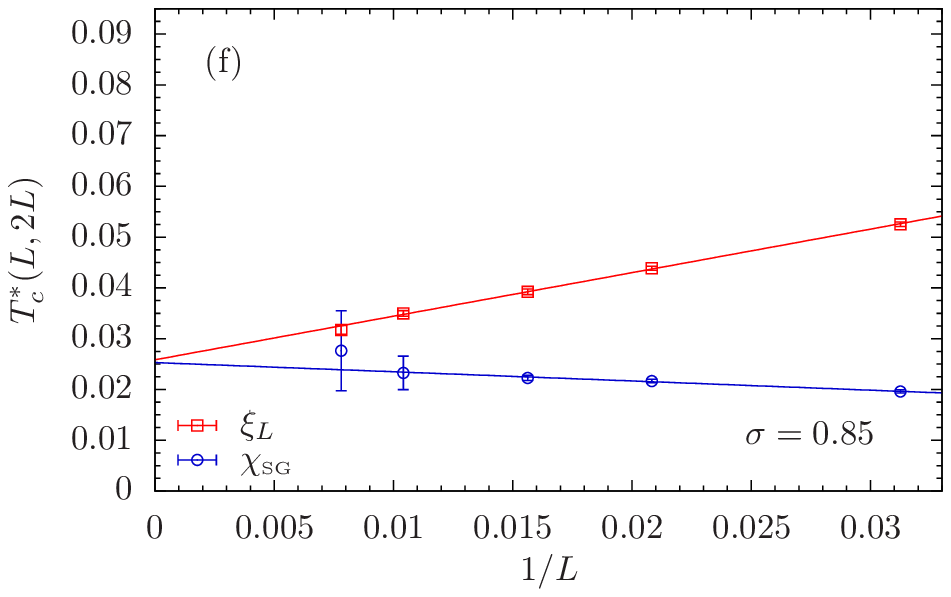}
\vspace*{-0.3cm}
\caption{(Color online)
Panels (a), (c) and (e) show the correlation length $\xi_L/L$
(inset: susceptibility $\chi_{\rm SG}/L^{2-\eta}$) as a function
of temperature $T$ for different system sizes $L$. Panel (a) shows
data for $\sigma = 0.60$ (mean-field regime) where a transition is
expected \cite{brangian:02c} [note that here $\nu = 1/(2\sigma -
1)$]. Panels (c) and (e) show data for $\sigma = 0.75$ and $\sigma =
0.85$, respectively, which correspond to a space dimension below
the upper critical dimension.  A transition for low yet finite
temperature is clearly visible. Panels (b), (d) and (f) show the
crossing temperatures $T_c^*(L,2L)$ of successive pairs of system
sizes for different exponents $\sigma$ [(b) $0.60$; (d) $0.75$; (f)
$0.85$]. The crossings for both $\xi_L/L$ and $\chi_{\rm SG}/L^{2 -
\eta}$ are well approximated by a linear behavior in $1/L$. Despite
small deviations between the estimates for both quantities, for all
$\sigma$ values studied $T_c(\sigma) > 0$. In particular, we estimate
$T_{c}(0.60) = 0.060(4)$, $T_{c}(0.75) = 0.040(3)$ and $T_{c}(0.85) =
0.025(3)$. Note that the data for $\sigma = 0.60$ show a deviation from
the linear behavior for the largest system sizes studies. However,
both data sets agree and therefore suggest that the thermodynamic
limit might have been reached.
}
\vspace{-.1cm}
\label{fig:crossings}
\end{figure*}

\section{Results}
\label{sec:results}

Our results are summarized in Fig.~\ref{fig:crossings}.
The main panels in the left column show data for the finite-size correlation
length as a function of temperature for (a) $\sigma = 0.60$, (c) $0.75$, 
and (e) $0.85$. The insets show the corresponding data
for the scaled dimensionless susceptibility.  In all cases data for
different system sizes cross, indicating the presence of a transition.
To better quantify the thermodynamic behavior, we show in the right
column the scaling of the crossing between successive system size
pairs $T^*(L,2L)$ as a function of $1/L$. The data can be well fit by
a linear function; the intercept with the vertical axis corresponding
to the thermodynamic limit. For all $\sigma$ studied we find finite
values for the thermodynamic glass transition.  These findings for the
long-range model with power-law interactions imply that the 10-state
mean-field Potts glass, for $d_{\rm u} < d < \infty$ space dimensions,
has a stable glass phase at finite temperatures.  In addition, our data
for $\sigma > 2/3$ indicate that short-range Potts glasses with a space
dimension below the upper critical dimension should also have a finite
transition temperature, albeit at very low~$T$ \cite{comment:rppg}.

Recently, Alvarez Ba\~nos {\em et al.}~\cite{alvarez:10-ea}
performed a thorough study of a three-dimensional Potts glass with
$p \le 6$, bimodal disorder and $J_0 = 0$. Their main result is
that $T_c$ decreases with an increasing number of states $p$ and
suggests that for $10$ states $T_c$ should be strongly suppressed, in
agreement with our results.  In addition, Alvarez Ba\~nos {\em et
al.}~\cite{alvarez:10-ea} claim that (1) only weak ferromagnetic
order is visible when $J_0 = 0$, (2) that the complexity of the
simulations is much higher when $J_0 = 0$, (3) that setting $J_0 =
-1$ could impact the presence of the glass transition, and (4) that
the transition could be first order.

\begin{figure}[h]
\vspace*{-0.2cm}
\centering
\includegraphics[width=1\columnwidth]{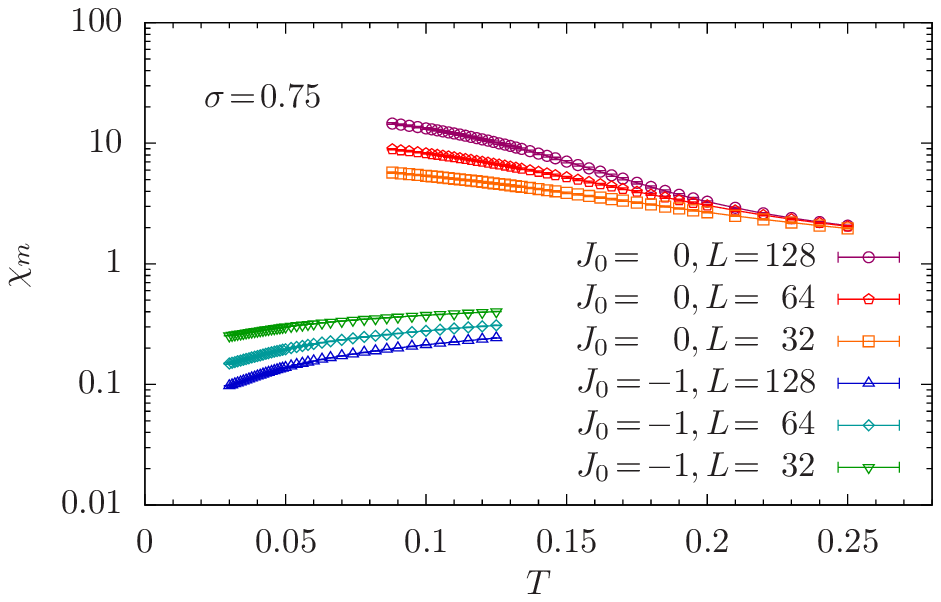}
\vspace*{-0.5cm}
\caption{(Color online)
Ferromagnetic susceptibility $\chi_m$ as a function of temperature $T$
for different system sizes. Ferromagnetic order is strongly suppressed
for $J_0 = -1$ in comparison to the $J_0 = 0$ case.
}
\vspace*{-0.1cm}
\label{fig:chim}
\end{figure}

We have examined these claims using the one-dimensional model
with Gaussian disorder and find that (1) ferromagnetic order
grows considerably when $J_0 = 0$ at low enough temperatures (see
Fig.~\ref{fig:chim}) and (2) the complexity of the simulations is
not affected by shifting the mean of the interactions.  With respect
to point (3), we do find, however, that the transition temperatures
are reduced by approximately a factor of $2$ -- $3$ when $J_0 =
-1$ in comparison to the simulations where $J_0 = 0$. Shifting the
mean of the interactions therefore only quantitatively impacts the
transition temperature.  Finally, (4), for the system sizes studied,
the distribution functions of the energy show no double-peak structure
that would be indicative of a first-order transition.

\section{Conclusions}
\label{sec:conlcusions}

Using a one-dimensional 10-state Potts glass with power law
interactions, we present evidence suggesting that short-range
finite-dimensional 10-state Potts glasses should exhibit a
finite-temperature transition for low enough temperatures and large
enough system sizes. Although corrections to scaling are large, we
estimate that for all $\sigma$ values studied $T_c(\sigma) > 0$. In
particular, we conservatively estimate $T_{c}(0.60) = 0.060(4)$,
$T_{c}(0.75) = 0.040(3)$, and $T_{c}(0.85) = 0.025(3)$. Larger
system sizes might show a different behavior, however, the presented
state-of-the-art simulations show strong evidence that short-range
10-state Potts glasses in high enough space dimensions should order.

\begin{acknowledgments} 
We thank A.~P.~Young for numerous discussions. H.G.K.~acknowledges
support from the SNF (Grant No.~PP002-114713). The authors acknowledge
ETH Zurich for CPU time on the Brutus cluster.
\end{acknowledgments}

\vspace*{-0.5cm}

\bibliography{refs,comments}

\begin{thebibliography}{31}
\bibitem[{com()}]{comment:rppg}
\bibinfo{note}{We have also studied the random-permutation Potts glass
  \cite{marinari:99a} (data not shown). However, equilibration turned out to be
  considerably harder and a reliable determination of the thermodynamic glass
  transition turned out to be impossible.}
\end{thebibliography}

\end{document}